\documentclass[a4paper,fleqn,usenatbib]{mnras}

\usepackage{graphicx}	
\usepackage{amsmath}	
\usepackage{amssymb}	
\usepackage{pdflscape}

\def\msun{\hbox{M$_\odot$}}

\title[vdBH open clusters]{A comprehensive photometric study of dynamically evolved small van den 
Bergh-Hagen open clusters}

\author[A.E. Piatti]{
Andr\'es E. Piatti$^{1,2}$\thanks{E-mail: andres@oac.unc.edu.ar}\\
$^{1}$Observatorio Astron\'omico, Universidad Nacional de C\'ordoba, Laprida 854, 5000, 
C\'ordoba, Argentina\\
$^{2}$Consejo Nacional de Investigaciones Cient\'{\i}ficas y T\'ecnicas, Av. Rivadavia 1917, 
C1033AAJ, Buenos Aires, Argentina\\
}

\date{Accepted XXX. Received YYY; in original form ZZZ}

\pubyear{2016}

\begin{document}
\label{firstpage}
\pagerange{\pageref{firstpage}--\pageref{lastpage}}
\maketitle

\begin{abstract}
We present results from Johnson $UBV$, Kron-Cousins $RI$ and Washington $CT_1T_2$
photometries for seven van den Bergh-Hagen (vdBH) open clusters, 
namely, vdBH\,1, 10, 31, 72, 87, 92, and 118.
The high-quality, multi-band photometric data sets were used to trace the cluster stellar
density radial profiles and to build colour-magnitude diagrams (CMDs) and colour-colour 
(CC) diagrams from which we estimated their structural parameters and fundamental astrophysical properties. The clusters in our sample
cover a wide age range, from $\sim$ 60 Myr up to 2.8 Gyr, are of relatively small
size ($\sim$ 1 $-$ 6 pc) and are placed at distances from the Sun which vary between
1.8 and 6.3 kpc, respectively. 
We also estimated lower limits for the cluster present-day masses as well as half-mass 
relaxation times ($t_r$). The resulting values in combination with the
structural parameter values suggest that the studied clusters are in
advanced stages of their internal dynamical evolution (age/$t_r$ $\sim$ 20 $-$ 320),
possibly in the typical phase of those tidally filled with mass segregation in their core 
regions. Compared to open clusters in the solar neighbourhood, the seven vdBH clusters are 
within more massive ($\sim$ 80 $-$ 380$\msun$), with higher concentration
parameter values ($c$ $\sim$ 0.75$-$1.15) and dynamically evolved  
ones. 
\end{abstract}




\section{Introduction}

\citet{vdbh75} performed a uniform survey over a $\sim$ 12$\degr$ strip
of the southern Milky Way extending from {\it l} $\approx$ 250$\degr$ and  {\it l} $\approx$ 
360$\degr$. They employed the Curtis-Schimidt telescope of the Cerro Tololo Interamerican
Observatory and a pair of blue and red filters. From that survey the authors
recognised 262 star clusters, 63 of which had not been previously catalogued.
For each idenfied object, they assessed the richness of stars on both plates as well as
the possible reality of being a genuine star cluster.

Up to date, less than 25 per cent of the van den Bergh-Hagen (vdBH) objects have some estimation 
of their fundamental properties (reddening, distance, age, etc). In general terms, 
according to the most updated version of the open cluster catalogue compiled by 
\citet[][version 3.5 as of January 2016]{detal02}, vdBH
clusters are mostly of relatively small size, with diameters smaller than $\sim$ 5 pc, although 
some few ones have diameters twice as big this value. On the other hand, although $\sim$ 60 per 
cent of them are located
inside a circle of 2 kpc in radius from the Sun, 
the remaining ones reach 
distances as large as $\sim$ 12 kpc. Indeed, nearly 15 per cent of the sample is located
at distances larger than 5 kpc. As for their ages, the vdBH clusters expand over an interesting
age regime, from those with some few Myr up to the older ones with more than 3 Gyr. At this point,
it appears interesting to estimate fundamental parameters of those overlooked vdBH objects,
particularly those located far away from the Sun, in order to improve our knowledge
of the Galactic open cluster system beyond the solar neighbourhood.

In this paper, we present a comprehensive photometric study of vdBH\,1, 10, 31, 72, 87, 92 and 118;
the last four clusters were discovered by \citet{vdbh75}. 
As far as we are aware, previous photometric studies were performed for vdBH\, 1 
(=Haffner\,7), 10 (= Ruprecht\,35) and 31 (= Ruprecht\,60)
\citep{moitinhoetal2006,vetal08,bb10,cetal13,giorgietal2015}. In Section 2 we describe the collection
and reduction of the available photometric data and their thorough treatment in order to build
extensive and reliable data sets. The cluster structural and fundamental parameters are derived
from star counts and colour-magnitude and colour-colour diagrams as described in Section 3.
The analysis of the results of the different astrophysical parameters obtained is carried out
in Section 4, where implications about the stage of their dynamical evolution are suggested.
Finally, Section 5 summarizes the main conclusion of this work.

\section{Data collection and reduction}

We make use of images obtained with the Johnson $UBV$, 
Kron-Cousins $RI$ and Washington $C$ filters, using a 4K$\times$4K CCD 
detector array (scale of 0.289$\arcsec$/pixel) attached to the 1.0-m 
telescope at the Cerro Tololo Inter-American Observatory (CTIO), Chile, in 2011
January 31--February 4 (CTIO program \#2011A-0114, image header information:
PI: Clari\'a, Observers: Clari\'a-Palma).
The nights were of photometric quality with a typical seeing of 1.1$\arcsec$.
The data sets used in this work were downloaded from the public website of the 
National Optical Astronomy Observatory (NOAO) Science Data Management (SDM) 
Archives\footnote{http://www.noao.edu/sdm/archives.php.}.The log of the observations 
is presented in Table~\ref{tab:table1}, where the
main astrometric and observational information is summarized.

\begin{table*}
\caption{Observations log of selected vdBH clusters.}
\label{tab:table1}
\begin{tabular}{@{}lccccccc}\hline
Cluster  &R.A.      &Dec.     &{\it l} &b       &  filter & exposure & airmass\\
         &(h m s)   &($\degr$ $\arcmin$ $\arcsec$)&(\degr)&(\degr)&  &  (sec)  & \\
\hline

vdBH\,1, Haffner\,7  &07 22 55.0&-29 30 00&242.6732&-06.8043& $U$ &60, 480    &1.01, 1.01      \\
         &          &         &        &        & $B$ &60, 360    &1.00, 1.00      \\
         &          &         &        &        & $V$ &20, 60, 200&1.00, 1.00, 1.00\\
         &          &         &        &        & $R$ &15, 120    &1.00, 1.00      \\
         &          &         &        &        & $I$ &10, 90     &1.00, 1.00      \\
         &          &         &        &        & $C$ &50, 480    &1.00, 1.00      \\

vdBH\,10, Ruprecht\,35 &07 46 12.7&-31 16 59&246.6622&-03.2517& $U$ &90, 480    &1.00, 1.00      \\
         &          &         &        &        & $B$ &60, 360    &1.00, 1.00      \\
         &          &         &        &        & $V$ &60, 200    &1.01, 1.01      \\
         &          &         &        &        & $R$ &15, 120    &1.01, 1.01      \\
         &          &         &        &        & $I$ &10, 90     &1.01, 1.01      \\
         &          &         &        &        & $C$ &80, 480    &1.00, 1.00      \\ 

vdBH\,31, Ruprecht\,60 &08 24 25.9&-47 12 00&264.0909&-05.5022& $U$ &150, 600   &1.07, 1.08      \\
         &          &         &        &        & $B$ &90, 400    &1.05, 1.06      \\
         &          &         &        &        & $V$ &60, 180    &1.05, 1.05      \\
         &          &         &        &        & $R$ &20, 120    &1.05, 1.05      \\
         &          &         &        &        & $I$ &15, 90     &1.05, 1.05      \\
         &          &         &        &        & $C$ &100, 540   &1.07, 1.08      \\ 

vdBH\,72 &09 31 22.8&-53 02 06&275.4908&-01.1708& $U$ &60, 480    &1.09, 1.09      \\
         &          &         &        &        & $B$ &60, 300    &1.09, 1.09      \\
         &          &         &        &        & $V$ &20, 60, 180&1.09, 1.09, 1.09\\
         &          &         &        &        & $R$ &20, 120    &1.09, 1.09      \\
         &          &         &        &        & $I$ &10, 90     &1.09, 1.09      \\
         &          &         &        &        & $C$ &50, 420    &1.09, 1.09      \\ 

vdBH\,87 &10 04 18.0&-55 26 00&280.7188&+00.0590& $U$ &80, 420    &1.11, 1.11      \\
         &          &         &        &        & $B$ &60, 240    &1.11, 1.11      \\
         &          &         &        &        & $V$ &60, 180    &1.12, 1.12      \\
         &          &         &        &        & $R$ &20, 120    &1.12, 1.12      \\
         &          &         &        &        & $I$ &15, 90     &1.13, 1.12      \\
         &          &         &        &        & $C$ &80, 360    &1.11, 1.11      \\

vdBH\,92  &10 18 54.0&-56 26 00&282.9677&+00.4072& $U$ &90, 540    &1.12, 1.12      \\
         &          &         &        &        & $B$ &60, 360    &1.14, 1.14      \\
         &          &         &        &        & $V$ &60, 200    &1.14, 1.15      \\
         &          &         &        &        & $R$ &60, 120    &1.15, 1.15      \\
         &          &         &        &        & $I$ &10, 90     &1.16, 1.16      \\
         &          &         &        &        & $C$ &80, 480    &1.12, 1.13      \\

vdBH\,118&11 22 30.0&-58 31 48&291.5288&+02.3588& $U$ &120        &1.19            \\
         &          &         &        &        & $B$ &90, 360    &1.16, 1.17      \\
         &          &         &        &        & $V$ &30, 240    &1.16, 1.16      \\
         &          &         &        &        & $R$ &60, 240    &1.15, 1.15      \\
         &          &         &        &        & $I$ &15, 180    &1.15, 1.15      \\
         &          &         &        &        & $C$ &90, 300    &1.18, 1.19      \\

\hline
\end{tabular}
\end{table*}

The observations were supplemented with series of bias, dome and 
sky flat exposures per filter during the observing nights to calibrate the 
CCD instrumental signature. 
The data reduction followed the procedures documented
by the CTIO Y4KCam\footnote{http://www.ctio.noao.edu/noao/content/y4kcam} team and utilized the {\sc quadred} package in IRAF\footnote{IRAF is distributed by the National 
Optical Astronomy Observatories, which is operated by the Association of 
Universities for Research in Astronomy, Inc., under contract with the National 
Science Foundation.}. We performed overscan, trimming, bias subtraction,
flattened all data images, etc., once the calibration frames were properly 
combined.

Nearly 150 independent magnitude measures of stars in the standard fields
SA\,98 and SA\,101 \citep{l92,g96} were also 
derived per filter for each night using the {\sc apphot} task within IRAF, in order 
to secure the transformation from the instrumental to the Johnson-Kron-Cousins 
$UBVRI$ and Washington $CT_1T_2$ standard systems. {Note that $T_2$ is 
$I_{KC}$ \citet{g96}.} The relationships between instrumental
and standard magnitudes were obtained by fitting the equations:

\begin{equation}
u = u_1 + V + (U-B) + u_2\times X_U + u_3\times (U-B),
\end{equation}

\begin{equation}
b = b_1 + V + (B-V) + b_2\times X_B + b_3\times (B-V),
\end{equation}

\begin{equation}
v = v_1 + V + v_2\times X_V + v_3\times (V-I),			
\end{equation}

\begin{equation}
r = r_1 + V - (V-R) + r_2\times X_R + r_3\times (V-R),
\end{equation}

\begin{equation}
i = i_1 + V - (V-I) + i_2\times X_I + i_3\times (V-I),
\end{equation}

\begin{equation}
c = c_1 + T_1 + (C-T_1) + c_2\times X_C + c_3\times (C-T_1),
\end{equation}

\begin{equation}
r = {t_1}_1 + T_1 + {t_1}_2\times X_{T_1} + {t_1}_3\times (C-T_1),
\end{equation}

\begin{equation}
t_2 = {t_2}_1 + T_1 - (T_1-T_2)  + {t_2}_2\times X_{T_1} + {t_2}_3\times (T_1-T_2),
\end{equation}

\noindent where $u_i$, $b_i$, $v_i$, $r_i$, $i_i$, $c_i$, ${t_1}_i$ and 
${t_2}_i$ ($i$ = 1, 2 and 3) 
are the fitted coefficients, and $X$ represents the effective airmass. Capital and 
lowercase letters 
represent standard and instrumental magnitudes, respectively.
Here, we use lower case $r$ for the $T_1$ filter because
we indeed used the $R(KC)$ filter as more efficient substitute of the Washington
$T_1$ filter, as shown by \citet{g96}.The $r$ magnitudes were thus transformed to
$T_1$ magnitudes, keeping in mind that the latter are not strickly the standard 
$T_1$ magnitudes, since some difference in the filter transmision curve could exit. 
We solved the transformation
equations with the {\sc fitparams} task in IRAF for each night, and found 
mean colour terms of 0.056 in $u$, 0.117 in $b$,
 -0.022 in $v$, -0.003 in $r$, -0.022 in $i$, -0.016 in $c$, -0.001 in $t_1$ and
0.016 in $t_2$, and extinction coefficients of 0.491 in $u$, 0.327 in $b$,
 0.093 in $v$, 0.095 in $r$, 0.056 in $i$, 0.514 in $c$, 0.096 in $t_1$ and
0.045 in $t_2$; the rms errors from the transformation to the standard system
are 0.071 in $u$, 0.054 in $b$, 0.050 in $v$, 0.028 in $r$, 0.031 in $i$, 0.043 in 
$c$, 0.037 in $t_1$ and 0.038 in $t_2$, respectively.  The latter can be the
result of the combination of several reasons, among them, the transmision curve of 
the filters used, the quantum eficiency of the CCD towards blue/near-IR wavelengths, 
slight weather variations during an observing night (sometime in some particular region 
of the sky), standard stars observed in some few moments during an observing night, etc.
Since we are making use of available public data, it is not easy to assess about which one
of these reasons could be affecting the observations more significantly.

\begin{figure}
	\includegraphics[width=\columnwidth]{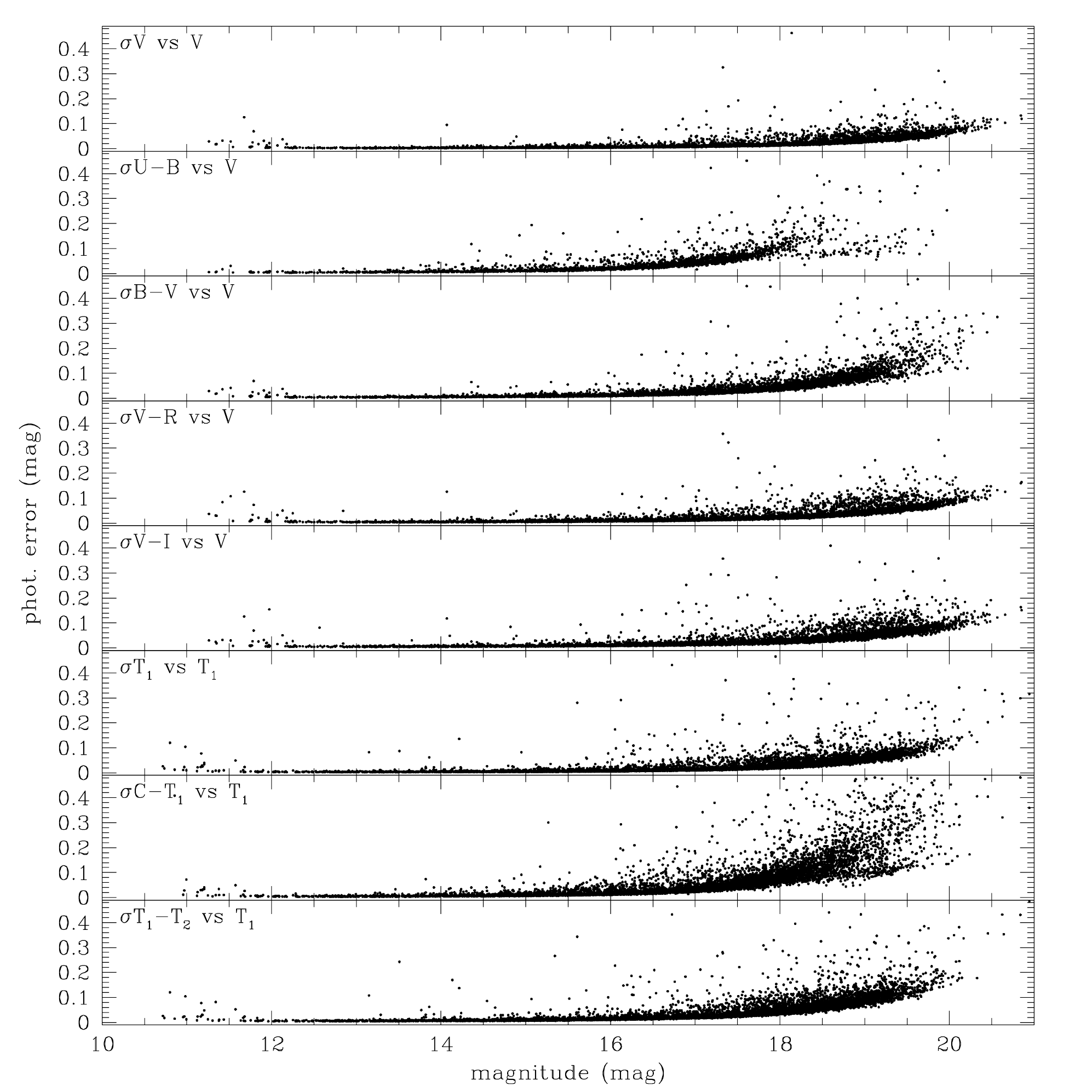}
    \caption{Photometric uncertainties of stars measured in the field of
vdBH\,72.}
    \label{fig:fig1}
\end{figure}

The stellar photometry was performed using the star-finding and point-spread-function (PSF) fitting 
routines in the {\sc daophot/allstar} suite of programs \citep{setal90}. For each image, 
a quadratically varying 
PSF was derived by fitting $\sim$ 200 stars, once the neighbours were eliminated using a preliminary PSF
derived from the brightest, least contaminated $\sim$ 60 stars. Both groups of PSF 
stars were interactively selected. We then used the {\sc allstar} program to apply the resulting PSF to the 
identified stellar objects and to create a subtracted image which was used to find and measure magnitudes of 
additional fainter stars. This procedure was repeated three times for each frame. 
After deriving the photometry for all detected objects in
each filter, a cut was made on the basis of the parameters
returned by DAOPHOT. Fig.~\ref{fig:fig1} illustrates the typical uncertainties in the 
derived photometry. Only objects with $\chi$ $<$2, photometric error less than 2$\sigma$ above the mean error at a given
magnitude, and $|$SHARP$|$ $<$ 0.5 were kept in each image, for which we also computed aperture corrections. 

We combined the individual $U,B,V,R,I$ photometric files using the stand-alone {\sc daomatch} 
and {\sc daomaster} programs\footnote{Kindly provided by P. Stetson. by
requesting that at least one colour can be computed during the matching of all the
photometric information for each star. Similarly, we gathered the $C,T_1,T_2$ photometric files.
Finally, we produced 2 or 3 independent $UBVRI$ and $CT_1T2$ data sets, depending on the
number of observation per filter available.}
We standardised the resulting individual data sets
from eqs. 1--8, then averaged the standard magnitudes and colours of each star
in the different data sets, and finally
cross-matched the averaged  $UBVRI$ and $CT_1T_2$ data sets to build one master table
per cluster field. The final information for each cluster field 
consists of a running number per star, its $x$ and $y$ coordinates, the mean $V$ 
magnitude, its rms error and the number of measurements, the colours $U-B$, $B-V$,
$V-R$, $V-I$ with their respective rms errors and number of measurements, the $T_1$
magnitude with its error and number of measurements, and the $C-T_1$ and $T_1-T_2$
colours with their respective rms errors and number of measurements.
Table~\ref{tab:table2} gives this information for vdBH\,1. Only a portion 
of this table is shown here for guidance regarding its form and content. The whole content 
of Table~\ref{tab:table2}, as well as those for the remaining cluster fields, is 
available in the online version of the journal.

\begin{table*}
\caption{$UBVRI$ and $CT_1T_2$ data of stars in the field of vdBH\,1.}
\label{tab:table2}
\tiny
\begin{tabular}{@{}lcccccccccc}\hline
Star & $x$ & $y$ & $V$ & $U-B$ & $B-V$ & $V-R$ & $V-I$ &
$T_1$ & $C-T1$ & $T_1-T_2$ \\
   & (pixel) & (pixel) & (mag)   & (mag)  & (mag)  & 
(mag)    & (mag)    & (mag)    & (mag)   & 
(mag)   \\\hline
 -- & --& --& -- & --& --& -- & --& --& -- & --  \\
   1755& 3321.011& 2042.219 &  16.890   0.006  2  &  0.150    0.029  1&    0.694    0.017  1 &   0.436    0.019  2&    0.766    0.029  2&   16.533    0.008  2 &   1.332    0.002  2&    0.364    0.022  2\\
   1756 &1641.079 &2042.294 &  17.099    0.012  1   & 0.014    0.034  1 &   0.691    0.020  1 &   0.440    0.017  1 &   0.330    0.063  1 &  16.717    0.025  2 &   1.233    0.003  2 &  -0.107    0.063  1\\
   1757  &797.882 &2043.852 &  16.673    0.021  2 &   0.147    0.031  1&    0.823    0.017  1&    0.503
0.024  2&    0.921    0.028  2 &  16.249    0.002  2 &   1.548    0.020  2 &   0.461    0.015  2\\
 -- & --& --& -- & --& --& -- & --& --& -- & -- \\
\hline
\end{tabular}
\end{table*}

\section{Cluster properties}

\subsection{Structural parameters}

\begin{figure}
	\includegraphics[width=\columnwidth]{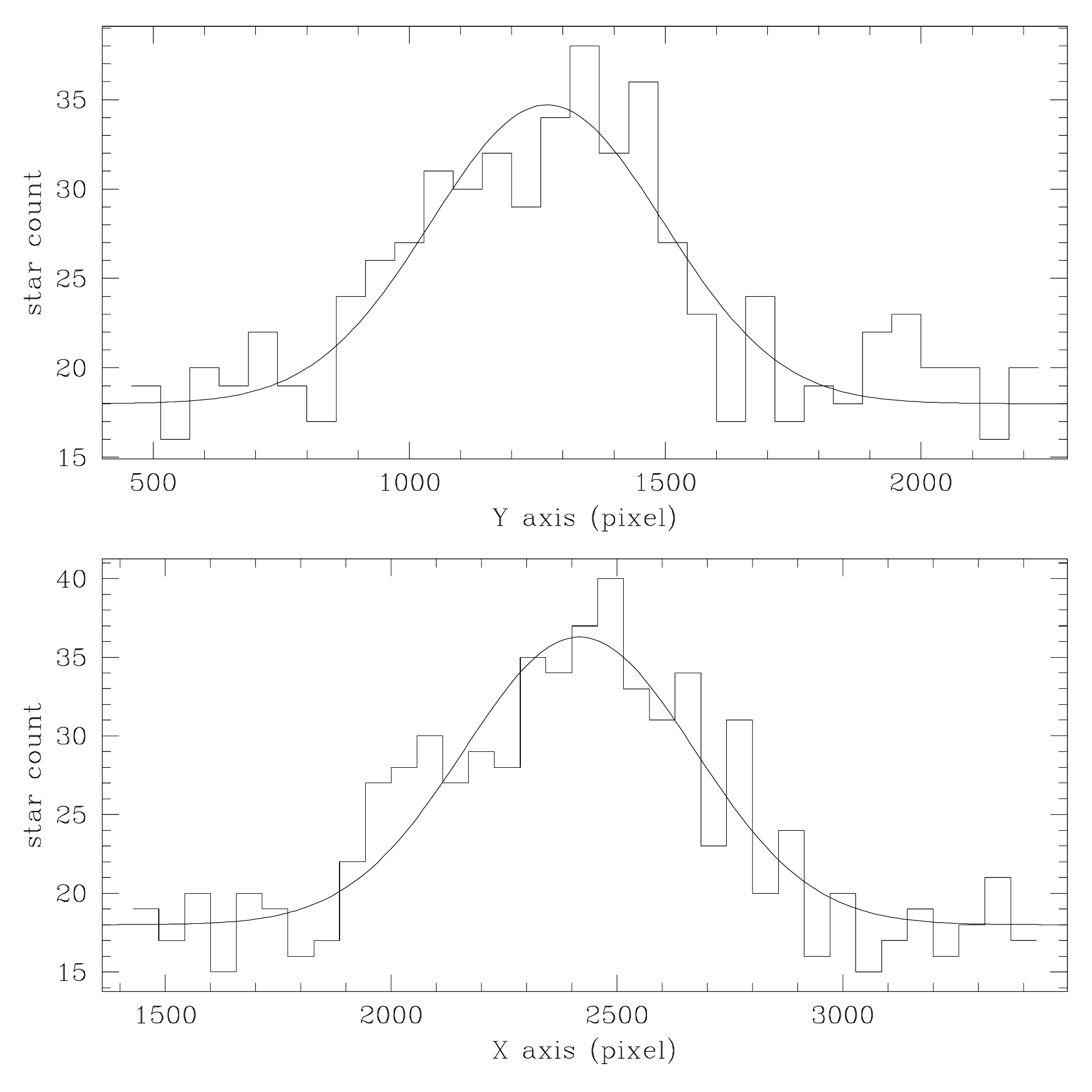}
    \caption{Star count along the $x$ and $y$ directions in the field of
vdBH\,72.}
    \label{fig:fig2}
\end{figure}

We determined the geometrical centres of the clusters in order to obtain their stellar density radial 
profiles.
The coordinates of the cluster centres and their estimated uncertainties were determined by fitting Gaussian 
distributions to the star counts in the $x$ and $y$ directions for each cluster. The fits of the Gaussians
 were 
performed using the {\sc ngaussfit} routine in the {\sc stsdas/iraf} package. We adopted a single Gaussian
 and fixed the constant 
to the corresponding background levels (i.e. stellar field densities assumed to be uniform) and the linear 
terms to zero. The centre of the Gaussian, its amplitude, and its $FWHM$ acted as variables. 
The number of stars projected along the $x$ and $y$ directions were counted within intervals 
of 20, 30, 40, 50 and 60 pixel wide, and the Gaussian fits repeated each time. Finally,
we averaged the five different Gaussian centres with a typical standard deviation
of $\pm$ 20 pixels ($\pm$  5.6$\arcsec$) in all cases. Fig.~\ref{fig:fig2}
illustrates the results of this procedure for vdBH\,72.

\begin{figure*}
	\includegraphics[width=\columnwidth]{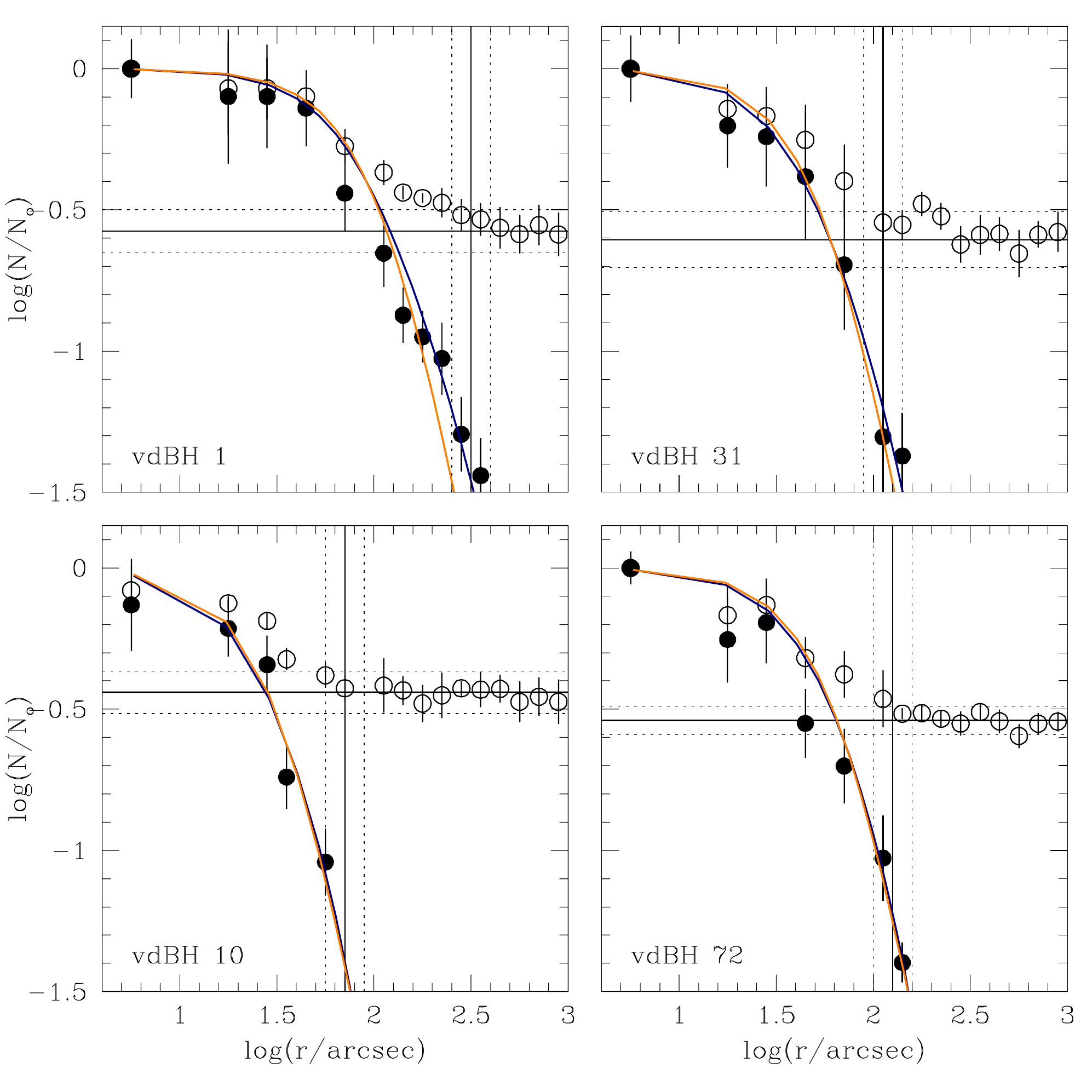}
        \includegraphics[width=\columnwidth]{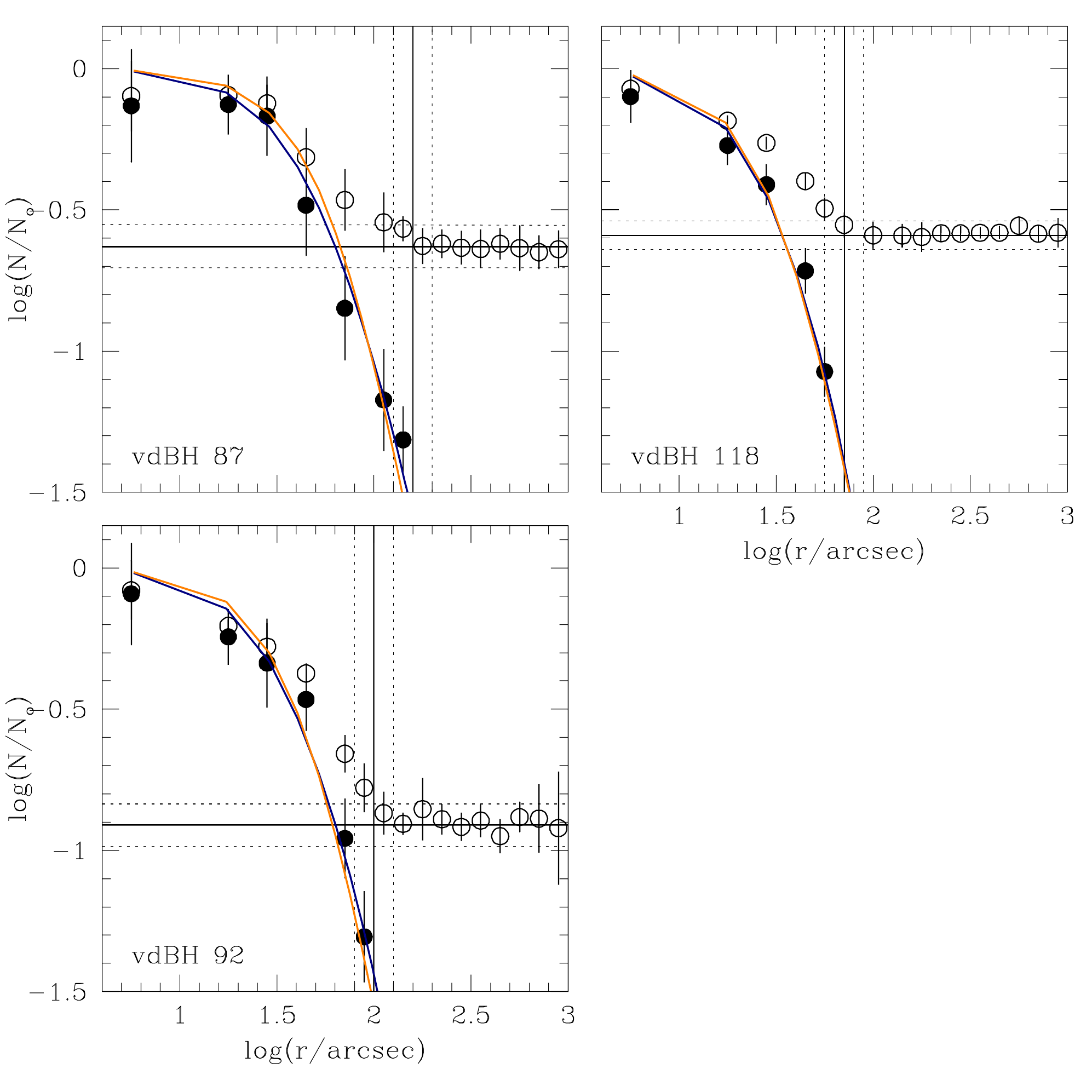}
    \caption{Stellar density profiles obtained from star counts. Open and filled circles 
refer to measured and
background subtracted density profiles, respectively. Blue and orange solid lines
depict the fitted King and Plummer curves, respectively.}
    \label{fig:fig3}
\end{figure*}

We also built stellar density profiles based on star counts previously performed within boxes of 40 pixels per side distributed 
throughout the whole field of each cluster. This box size allowed us to sample  
the stellar spatial distribution statistically. 
Thus, the number of stars per unit area at a given radius $r$ can be directly calculated through 
the expression:

\begin{equation}
(n_{r+20} - n_{r-20})/(m_{r+20} - m_{r-20}),
\end{equation}

\noindent where $n_r$ and $m_r$ represent the number of stars and boxes, respectively,  included in a circle of radius $r$. We note that this method does not necessarily require a complete circle of radius $r$ within the
 observed 
field to estimate the mean stellar density at that distance. This is an important consideration since having a stellar 
density profile that extends far away from the cluster centre allows us to estimate the background level with 
high precision. This is necessary in order to derive the cluster radius ($r_{cls}$). The resulting density profiles 
expressed as number of stars per arcsec$^2$ are shown in  Fig.~\ref{fig:fig3}.
In the figure, we 
represent the constructed and
background subtracted density profiles with open and filled circles, respectively.
Errorbars represent rms errors, to which we added the mean error of the
background star count to the background subtracted density profile. The background level
and the cluster radius are indicated by solid horizontal
and vertical lines, respectively; their uncertainties are in dotted lines.

The background corrected density profiles were fitted using a \citet{king62}'s model 
through the expression :
\begin{equation}
 N \varpropto ({\frac{1}{\sqrt{1+(r/r_c)^2}} - \frac{1}{\sqrt{1 + (r_t/r_c)^2}}})^2
\end{equation}

\noindent where 
$r_c$ and $r_t$ are the core and tidal radii, respectively (see Table~\ref{tab:table9} and Fig.~\ref{fig:fig3}). We used a grid of $r_c$ and $r_t$ values spanning the whole range of radii 
\citep{piskunovetal2007} and minimised $\chi$$^2$. 
The values derived for $r_c$ and 
$r_t$ from the fit are listed in Table~\ref{tab:table9}, while the respective King's curves are plotted with blue solid lines in Fig.~\ref{fig:fig3}.
As can be seen, the King profiles satisfactorily reproduce the whole cluster extensions.
Nevertheless, in order to get independent estimates of the star cluster half-mass radii, we
fitted Plummer's profiles using the expression:
\begin{equation}
N \varpropto \frac{1}{(1+(r/a)^2)^2} 
\end{equation}

\noindent where $a$ is the Plummer's radius, which is related to the half-mass radius ($r_h$) by the relation $r_h$ $\sim$ 1.3$a$. 
The resulting $r_h$ values are listed in Table~\ref{tab:table9} and the corresponding Plummer's curves are drawn with orange solid lines in Fig.~\ref{fig:fig3}.

\setcounter{table}{8}
\begin{table*}
\centering
\caption{Fundamental parameters of vdBH open clusters.}
\label{tab:table9}
\begin{tabular}{@{}lcccccccccccc}\hline\hline
Star cluster &   $E(B-V)$ & $(m-M)_o$ & d  &  $r_c$ & $r_h$ &  $r_{cls}$ &  $r_t$  & $\log(t)$ & Z$_{\rm iso}$ &$M_{cls}$ & $t_r$\\
             &  (mag)  & (mag)  & (kpc)    &   (pc) &  (pc) &   (pc)     &  (pc)  &  &  & ($\msun$) & (Myr) \\\hline
vdBH\,1      &   0.10$\pm$0.03  & 13.10$\pm$0.20  & 4.2$^{+0.5}_{-0.4}$ & 1.6$\pm$0.2  & 3.0$\pm$0.3 & 6.4$^{+1.6}_{-1.3}$&22.4$\pm$2.0 & 9.10 & 0.0152 & 378$\pm$32 & 42.0\\
vdBH\,10     &    0.60$\pm$0.10  & 12.50$\pm$0.20  & 3.2$^{+0.3}_{-0.2}$ & 0.4$\pm$0.1  & 0.7$\pm$0.2 & 1.1$^{+0.3}_{-0.2}$ & 2.3$\pm$0.2 & 7.80 & 0.0152 &120$\pm$10 & 3.0\\
vdBH\,31     &   0.30$\pm$0.05  & 12.30$\pm$0.20  & 2.9$^{+0.3}_{-0.2}$ & 0.6$\pm$0.1  & 1.1$\pm$0.2 & 1.6$^{+0.4}_{-0.3}$ & 4.9$\pm$1.4 & 9.10 & 0.0152 & 133$\pm$12 & 7.0\\
vdBH\,72     &   0.85$\pm$0.10  & 12.60$\pm$0.30  & 3.3$^{+0.5}_{-0.4}$ & 0.8$\pm$0.2  & 1.4$\pm$0.2 & 2.0$^{+0.5}_{-0.4}$ & 4.8$\pm$0.8 & 8.55 & 0.0152 &301$\pm$26 & 13.0\\
vdBH\,87     &    0.45$\pm$0.05  & 11.30$\pm$0.20  & 1.8$^{+0.2}_{-0.1}$ & 0.3$\pm$0.1  & 0.6$\pm$0.1 & 1.4$^{+0.4}_{-0.3}$& 3.5$\pm$0.9 &8.50 & 0.0152 & 132$\pm$12 & 3.0\\
vdBH\,92     &    0.35$\pm$0.05  & 11.50$\pm$0.20  & 2.0$^{+0.2}_{-0.1}$ & 0.3$\pm$0.1 & 0.6$\pm$0.2 & 1.0$^{+0.3}_{-0.2}$& 2.4$\pm$0.5 &8.20 & 0.0152 & 84$\pm$8 & 3.0\\
vdBH\,118    &    0.35$\pm$0.05  & 14.00$\pm$0.25  & 6.3$^{+0.8}_{-0.7}$ & 0.8$\pm$0.2  & 1.4$\pm$0.3 & 2.2$^{+0.5}_{-0.4}$ & 4.6$\pm$0.9 &9.45 & 0.0110 & 113$\pm$10 & 9.0\\

\hline
\end{tabular}

\noindent Note: to convert 1 arcsec to pc, we use the following expression,10$\times$10$^{(m-M)_o/5}$sin(1/3600), 
where $(m-M)_o$ is the true distance modulus.

\end{table*}

\subsection{Colour-magnitude diagram analysis}

\subsubsection{Cleaning the cluster colour-magnitude diagrams}

We used the mean cluster radii to extract the cluster colour-magnitude
diagrams (CMDs). They account for the luminosity function, colour distribution and
stellar density of the stars distributed along the cluster line of sights, so that we 
statistically cleaned the CMDs before using them to estimate the
cluster fundamental parameters.

We employed the cleaning procedure developed by \citet[see their Fig. 12]{pb12}. 
The method compares the extracted cluster CMD to
distinct CMDs composed of stars located reasonably far from the
object, but not too far so as to risk losing the local field-star
signature in terms of stellar density, luminosity function and/or
colour distribution. Here we chose four field regions, each one designed to cover an
equal area as that of the cluster, and placed around the cluster. 
Note that the four selected fields could not
adequately represent the fore/background of the cluster if the extinction varies
significantly accross the field of view.

Comparisons of field and cluster CMDs have long been done by comparing
the numbers of stars counted in boxes distributed in a similar manner
throughout both CMDs. However, since some parts of the CMD are more
densely populated than others, counting the numbers of stars within
boxes of a fixed size is not universally efficient. For instance, to
deal with stochastic effects at relatively bright magnitudes (e.g.,
fluctuations in the numbers of bright stars), larger boxes are
required, while populous CMD regions can be characterized using
smaller boxes. Thus, the use of boxes of different sizes distributed in
the same manner throughout both CMDs leads to a more meaningful
comparison of the numbers of stars in different CMD regions. 
Precisely, the procedure of \citet{pb12} carries out the comparison between
field-star and cluster CMDs by using boxes which vary their
sizes from one place to another throughout the CMD and are centred on the
positions of every star found in the field-star CMD.

By starting with reasonably large boxes -- typically 
($\Delta$(magnitude),$\Delta$(colour)) = (1.00, 0.50) mag -- centred on each
star in the four field CMDs and by subsequently reducing their sizes
until they reach the stars closest to the boxes' centres in magnitude and colour,
separately, we defined boxes which result in use of larger areas in
field CMD regions containing a small number of stars, and vice versa.
Note that the definition of the position and size of each box involves
two field stars, one at the centre of the box and another -the closest one to 
box centre - placed on the boundary of that box.
Next, we plotted all these boxes
for each field CMD on the cluster CMD and subtracted the star located
closest to each box centre. Since we repeated this task for each of the four 
field CMD box
samples, we could assign a membership probability to each star in the
cluster CMD. This was done by counting the number of times a star
remained unsubtracted in the four cleaned cluster CMDs and by
subsequently dividing this number by four. Thus, we distinguished
field populations projected on to the cluster area, i.e., those stars
with a probability $P$ $\le$ 25\%; stars that could equally likely be
associated with either the field or the object of interest ($P$ =
50\%); and stars that are predominantly found in the cleaned cluster
CMDs ($P \ge$ 75\%) rather than in the field-star CMDs. 
Statistically speaking, a certain amount of cleaning
residuals is expected, which depends on the degree of variability of the stellar density,
luminosity function and colour distribution of the field stars.

\begin{figure*}
	\includegraphics[width=\textwidth]{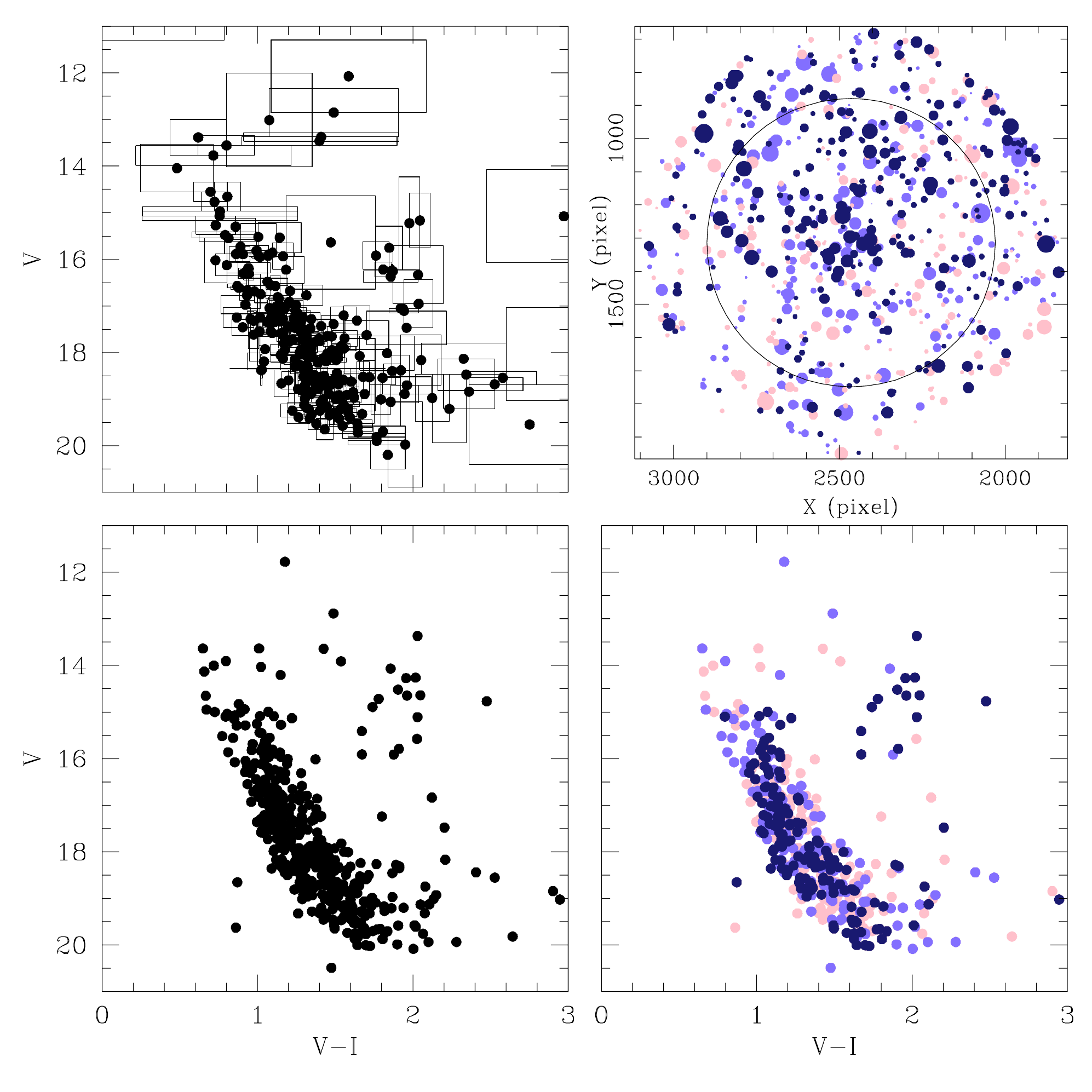}
    \caption{CMDs for stars in the field of vdBH\,72: the observed CMD
  composed of the stars distributed within the cluster radius (bottom
  left-hand panel); a field CMD for a circular region placed far from the cluster and
 with a size equal to the cluster area (top left-hand panel), and the corresponding
defined set of boxes overplotted; the cleaned cluster CMD (bottom right-hand panel). 
Colour-scaled symbols represent stars that statistically
  belong to the field ($P \le$ 25\%, pink), stars that might belong to
  either the field or the cluster ($P =$ 50\%, light blue), and stars
  that predominantly populate the cluster region ($P \ge$ 75\%, dark
  blue). The schematic finding chart for the cluster area is
  shown in the top right-hand panel. The black circle represents
  the adopted cluster radius. Symbols are as in the bottom right-hand
  panel, with sizes proportional to the stellar brightnesses. North is
  up; East is to the left.}
    \label{fig:fig4}
\end{figure*}

Fig.~\ref{fig:fig4}
illustrates the performance of the cleaning procedure in the field of vdBH\,72, where we
plotted three different ($V$,$V-I$) CMDs: a single field-star CMD (top left-hand panel) 
for a circular region with an area equal to that of the cluster; that for the stars 
located within the cluster radius (bottom left-hand panel); and the cleaned cluster
CMD (bottom right-hand panel) for stars with $P \le$ 25\% (pink), $P =$ 50\% (light blue)
and $P \le$ 75\% (dark blue). In the field-star CMD we 
overplotted the boxes generated by the cleaning procedure. A schematic finding chart 
with a circle of radius equal to the cluster radius is shown in the top 
right-hand panel.

\subsubsection{Fundamental cluster parameters}

Figures~\ref{fig:fig5} to \ref{fig:fig11} show the whole set of CMDs and colour-colour (CC) diagrams
for the cluster sample that can be exploited from the present extensive multi-band photometry. 
They include every magnitude and colour measurements of stars located within the
respective cluster radii (see Table~\ref{tab:table9}). We have also incorporated to the figures
the statistical photometric memberships obtained in Sect. 3.2.1 by
distinguishing stars with different colour symbols as in Fig.~\ref{fig:fig4}. At first glance, the cleaned
cluster CMDs (stars with $P \ge$ 75\%) resemble those moderately young to intermediate-age, projected on to
star fields with different levels of crowdness. 

\citet{pp15} introduced a new age-metallicity diagnostic diagram for the Washington photometric
system, $\delta$$T_1$ versus
$\delta$$C$ - $\delta$$T_1$, which has shown the ability of unambiguously providing 
age and metallicity estimates,  simultaneously.  $\delta$$C$ and $\delta$$T_1$ are the respective
magnitude differences between the giant branch clump and the main sequence turnoff (MSTO).
The new procedure allows to derive ages from 1 up to 13 Gyr and metallicities [Fe/H] from -2.0 up 
to +0.5 dex, and is independent of the cluster reddening and distance modulus. We used here that 
procedure to estimate the age and metallicity of three clusters (vdBH\,1, 31 and 118) whose cleaned
CMDs show a handful of red clump (RC) stars (see Figs.~\ref{fig:fig5}, \ref{fig:fig7} and \ref{fig:fig11}), besides their MSTOs.  We used 
the cleaned cluster CMDs  to 
measure $C$ and $T_1$ magnitudes at the MSTO and RC, then computed 
$\delta$$C$ and $\delta$$T_1$ and entered into the age-metallicity diagnostic diagram to estimate 
cluster ages and metallicities. The resulting $\delta$$T_1$  and $\delta$$C$ - $\delta$$T_1$ values 
with their uncertainties are drawn in 
Fig.~\ref{fig:fig12}, where we have traced iso-age lines and marked iso-abundance positions using
colour-coded lines and filled circles, respectively. From this figure we estimated by interpolation
ages of 
1.3$\pm$0.2 Gyr,  1.5$\pm$0.2 Gyr and 2.9$\pm$0.4 Gyr for vdBH\,1, 31 and 118, respectively.
As for the mean metallicities, although rather more uncertain than the ages ($\sigma$[Fe/H]= 0.25 dex), 
the clusters appear to 
be of solar or slightly subsolar metal content. 

For the remaining clusters in our sample (vdBH\,10, 
72, 87 and 92) -- whose CMDs and CC diagrams resemble those of young clusters 
(see Figs.~\ref{fig:fig6}, \ref{fig:fig8}, \ref{fig:fig9} and \ref{fig:fig10}) -- 
we also adopted a solar metal content (see last column of Table~\ref{tab:table9}). Note that, by 
considering the whole metallicity 
range of the Milky Way open
clusters \citep[see, e.g.][]{paunzeretal2010,hetal14} and by using the theoretical isochrones of 
\citet{betal12},
the differences at the zero age main sequence (ZAMS) in $V-I$ and $T_1-T_2$ colours is
smaller than $\sim$ 0.08 and 0.04 mag, respectively. This result implies that
negligible differences between the ZAMSs for the cluster metallicity and that of solar
metal content would appear, keeping in mind the intrinsic 
spread of the stars in the $V$ vs $V-I$ and $T_1$ vs $T_1-T_2$ CMDs. 

The availability of six CMDs and three different CC diagrams covering wavelengths from
the blue up to the near-infrarred
allowed us to derive reliable ages, reddenings and distances for the studied clusters. 
Particularly noticeable in the $V$ vs $U-B$ CMD, but applicable to every CMD, the shape
of the main sequence (MS), its curvatures (those  less and more pronounced), the relative 
distance between the RC and the MSTO in magnitude and colour separately, among others, 
are features tightly 
related to the cluster age, regardless their reddenings and distances. For this reason, we 
started by selecting theoretical isochrones \citep{betal12} with the adopted cluster metallicities 
in order to chose those which best match the clusters' features in the CMDs. From our first
choices, we derived the cluster reddenings by shifting those isochrones in the three CC diagrams
following the reddening vectors until their bluest points coincided with the observed ones.
Note that this requirement allowed us to use the $V-R$ vs $R-I$ CC diagram as well, even though the reddening vector runs almost parallell to the cluster sequence.
Finally, the mean $E(B-V)$ colour excesses were used to properly shift the chosen isochrones
in the CMDs in order to derive the cluster true distance modulii by shifting the isochrones 
along the magnitude axes. 

We iterated this procedure whenever refinements in the cluster ages
were necessary. Nevertheless, we
found that isochrones bracketing the initial age choice  by $\Delta$
log($t$ yr$^{-1}$) = $\pm$0.10 represent the overall age uncertainties
owing to the observed dispersion in the cluster CMDs and CC diagrams. Although in some
cases the age dispersion is smaller than $\Delta$ log($t$ yr$^{-1}$) =
0.10, we prefer to keep the former value as an upper limit to our error
budget. In order to enter the isochrones into the CMDs and CC diagrams we used the following
ratios: $E(U-B)$/$E(B-V)$ = 0.72 + 0.05$\times$$E(B-V)$ \citep{hj56}; $E(V-R)$/$E(B-V)$ = 0.65,
$E(V-I)$/$E(B-V)$ = 1.25, $A_{V}$/$E(B-V)$ = 3.1 \citep{cetal89}; $E(C-T_1)$/$E(B-V)$ = 1.97,
$E(T_1-T_2)$/$E(B-V)$ = 0.692, $A_{T_1}$/$E(B-V)$ = 2.62 \citep{g96}.
The adopted best matched isochrones are overplotted on Figs.~\ref{fig:fig5} to \ref{fig:fig11}, 
while the resulting values with their errors for the cluster reddenings, true distance 
modulii and ages are listed in Table~\ref{tab:table9}. 

The masses of the clusters in our sample were derived by summing the individual masses of 
stars with membership probabilities $P \ge$ 75\%. The latter were obtained by interpolation 
in the theoretical isochrones traced in Figs.~\ref{fig:fig5} to \ref{fig:fig11} from the 
observed $V$ magnitude of each star, properly corrected by reddening and distance modulus. 
We estimate the uncertainty in the mass to be  $\sigma(\log(M_{cls}/\msun))$ $\sim$ 0.2 
dex. Note that this error comes from propagation of the $V$ magnitude errors in the
mass distribution along the theoretical isochrones. It does not reflect the
deviation of the cluster mass computed from stars with $P \ge$ 75\%
from the actual total cluster mass. Nevertheless, at first glance, the appearance of 
the cluster CMDs and CC diagrams ($P \ge$ 75\%) do not seem to significantly differ
from those including any other observed stars placed along the adopted isochrones
with $P <$ 75\%, thought to be cluster stars.
At the same time, the computed cluster masses include some unavoidable interlopers,
which mitigate the loss of some cluster stars. In the case of vdBH\,118, the 
derived mass should be considered as a very lower limit, since its CMDs  
barely reach the cluster MSTO. For the remaining clusters, our photometry reach
well below the fainter MS cluster stars.
Using the resulting masses and the half-mass radii $r_h$ of Table~\ref{tab:table9}, we
computed the half-mass relaxation times using the equation \citep{sh71}:

\begin{equation}
t_r = \frac{8.9\times 10^5 M_{cls}^{1/2} r_h^{3/2}}{\bar{m} log_{10}(0.4M_{cls}/\bar{m})}
,\end{equation}

\noindent where $M_{cls}$ is the cluster mass and $\bar{m}$ is the mean mass of 
the cluster stars. 
The derived masses and relaxation times are listed in Table~\ref{tab:table9}.
If we considered non-oberved stars with masses between 1 and 0.5 $\msun$ and the 
Salpeter's mass function, the relaxation times would increase in $\sim$ 10 per cent.

\setcounter{figure}{11}
\begin{figure} 
   \includegraphics[width=\columnwidth]{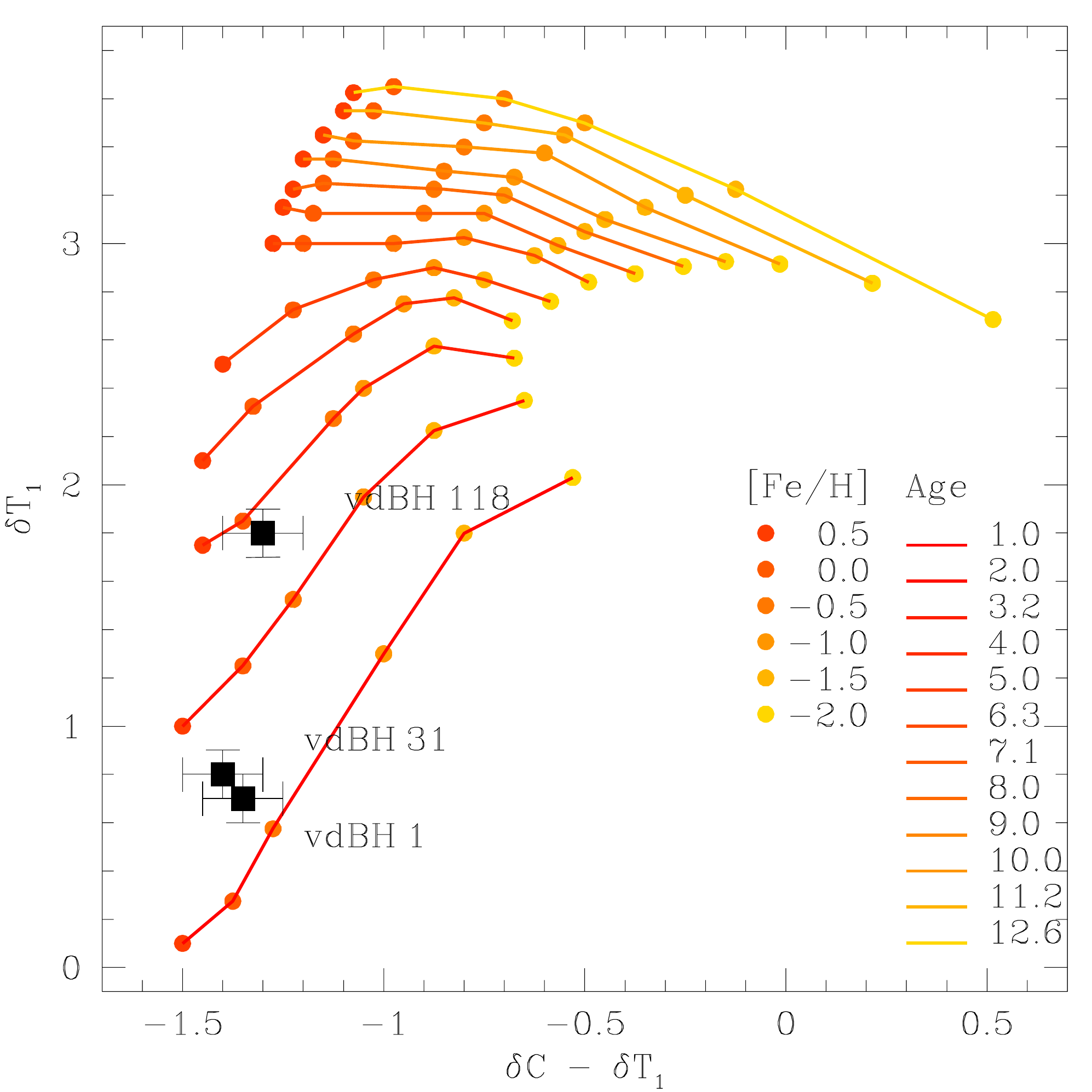}
\caption{$\delta$$T_1$ versus $\delta$$C$ - $\delta$$T_1$ diagram with iso-age lines and 
iso-metallicity locii. Metallicity and age labels are given in dex and Gyr, respectively.}
\label{fig:fig12}
\end{figure}

\section{results and discussion}

vdBH\,10 (= Ruprecht\,35) and 31 (= Ruprecht\,60) have age estimates of 400$\pm$100 Myr derived from Two-Micron All-Sky
Survey (2MASS)\footnote{The 2MASS, All Sky data release \citep{skrutskieetal2006} – http://www.
ipac.caltech.edu/2mass/releases/allsky/} photometry \citep{bb10}, which clearly differs from
our values (see Table~\ref{tab:table9}). By inspecting their 2MASS $J$ vs 
$J-H$ CMDs (Figures 4 and 5 in \citet{bb10}), and considering the relationship $M_J$ vs $M_V$ 
for the cluster ages computed by \citet{betal12} and our cluster distances,
we found that the faintest reached $J$ magnitude ($\approx$
15.5 mag) corresponds to $V$ $\approx$ 16.3 mag and 16.6 mag for vdBH\,10 and 31, respectively.
This suggests that the used 2MASS photometry is shallower than the present one.
On the other hand, we  speculate with the possibility that the field-star cleaning procedure
applied by \citet{bb10} have left significant residuals \citep[see][]{pb12,hanetal2016}; thus leading them 
to derive  ages which reflect the composite stellar population along the cluster line-of-sights.
Curiously, both clusters have nearly similar angular Galactic coordinates. Additionally, the 2MASS
CMDs were extracted using sky areas much larger than those embracing the clusters, so that
field-stars can quantitatively prevail over cluster stars. For these reasons, we are confidence
of the present multi-band photometry and resulting cluster parameters. 

The distance modulus and age of vdBH\,10 (= Ruprecht\,35) were previously estimated by \citet{moitinhoetal2006} and 
soon after used by \citet{vetal08} in an updated study of the spiral structure of the outer Galactic disc.
\citet{moitinhoetal2006} estimated a distance from the Sun of 5.32 kpc and an age of 70 Myr. While the cluster age is in
excellent agreement with the present value, the cluster distance differs significantly. This could be due
either to a photometric zero point offset or to the use of field star contaminated CMD diagrams in the analysis
carried out by \citet{moitinhoetal2006}. Both sources of error have been checked within our photometric data set.
On the one hand, we used six different CMDs involving $V$ and $T_1$ magnitudes, as well as $U-B$, $B-V$, $V-R$, $R-I$, $C-T_1$
and $T_1-T_2$ colours. On the other hand, we have had particular care in cleaning the cluster CMDs from field
star contamination (Sect. 3.2.1). As fas as we are aware, we are confident of the realibility of the present
cluster distance.

vdBH\,31 (= Ruprecht\,60) was
more recently studied by \citet{cetal13} using $UBVI$ photometry. The authors mentioned that
their photometry does not cover the object completely and that they did not perform any
cleaning of field stars in the cluster CMD.  They estimated an age of $\sim$1.5 Gyr,
a metallicity of [Fe/H] $\sim$ -0.5 dex, an $E(B-V)$ colour excess of 0.13$\pm$0.10 mag
and a distance from the Sun of 4.5 kpc. Their age, reddening and distance are
in very good agreement with our values. As for the remarkable low metal abundance, our 
metallicity sensitive Washington $C$ photometry do not support such a low value.  \citet{cetal13} could have
included field stars in their CMD which, alongside the fewer cluster stars measured, 
could lead to such a low value. 

Finally, \citet{giorgietal2015} studied vdBH\,31 (= Ruprecht\,60) also from
$UBVI$ photometry. They derived  $E(B-V)$ = 0.37$\pm$0.05 mag, a distance from the Sun of
4.2$\pm$0.2 kpc and an age of 0.8-1.0 Gyr. While the estimated reddening is in faily good 
agreeement with our value, they derived a slightly younger age which in turn led to estimate
a larger distance. We are confident of our age value which comes from a CMD analysis as
well as from the Washington age-metallicity diagnostic diagram analysis.

\begin{figure} 
   \includegraphics[width=\columnwidth]{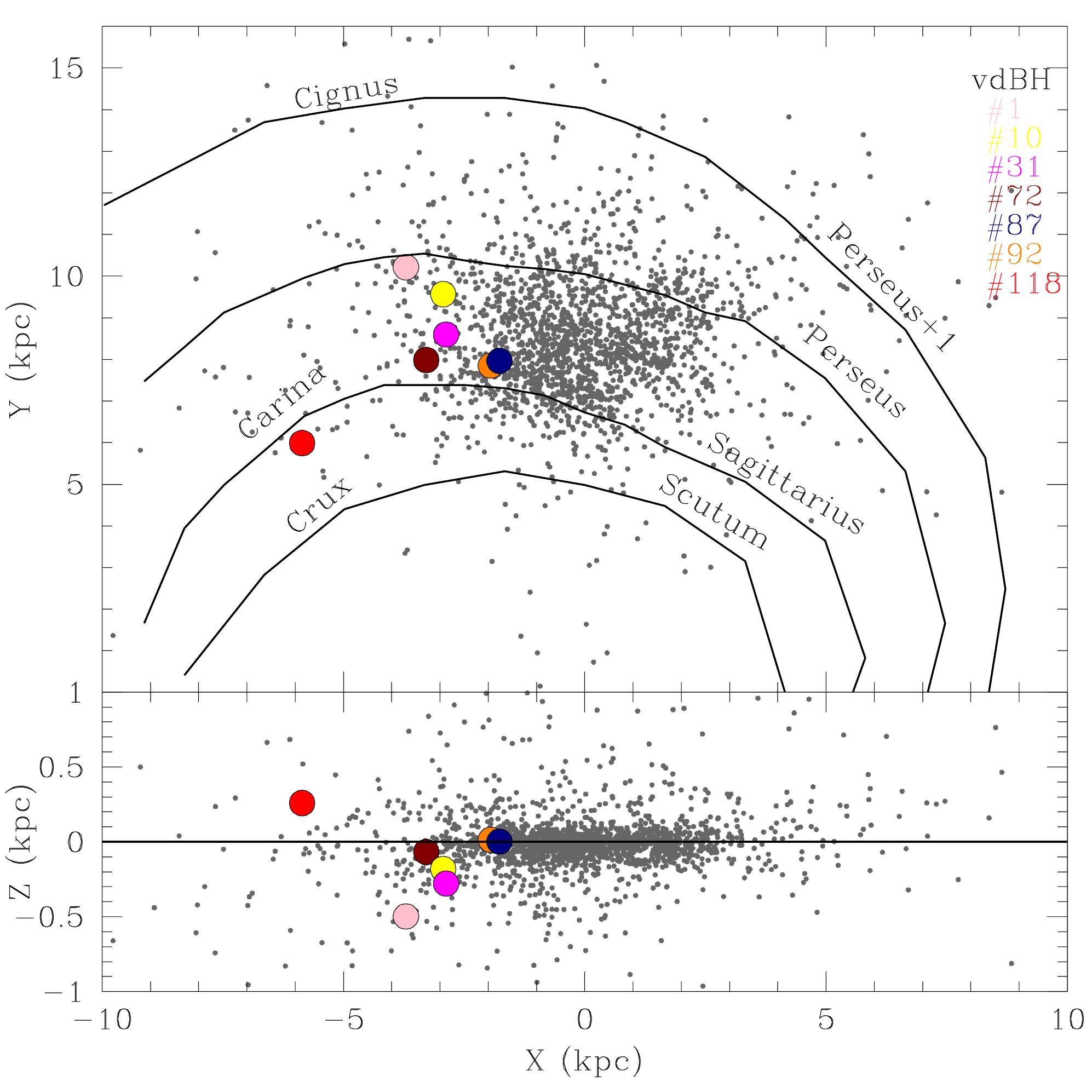}
\caption{Galactic spatial distribution of the studied clusters. Open clusters from the
catalogue of \citet[][version 3.5 as of January 2016]{detal02} are drawn with gray dots, while
the schematic positions of spiral arms \citep{ds2001,moitinhoetal2006} are traced with black 
solid lines.}
\label{fig:fig13}
\end{figure}

We computed Galactic coordinates using the derived cluster heliocentric distances, their angular
Galactic coordinates and a Galactocentric distance of the Sun of R$_{GC_\odot}$ = 8.3 kpc 
\citep[][and references therein]{hh2014}. The resulting spatial distribution is depicted
in Fig.~\ref{fig:fig13}, where we added for comparison purposes the 2167 open clusters 
catalogued by 
\citet[][version 3.5 as of January 2016]{detal02} and the schematic positions of the spiral
arms \citep{ds2001,moitinhoetal2006}. The studied clusters are mostly
located between the Carina and Perseus arms, with the sole exception of vdBH\,118, which lies
between the Carina and Crux spiral arms and is one of the most distant
open clusters from the Sun in that direction as well. In general terms, the 
seven vdBH clusters are distributed outside the circle around the Sun 
(d $\sim$ 2.0 kpc) where the catalogued clusters are mostly concentrated.

The structural parameters estimated from the stellar density radial profiles in combination with
the fundamental properties derived from the analysis of CMDs and CC diagrams allow us to 
investigate the internal dynamical state of the clusters. Here we consider the
effect of the internal dynamical evolution (two-body relaxation, mass segregation, etc). 
As for the Galactic tidal field,  bearing in mind the
Galactocentric distances of the studied clusters (R$_{GC}$ $\approx$ 8.1$-$10.9
kpc, with a mean value of 9$\pm$1 kpc) and their ages (see Table~\ref{tab:table9}),
the differences in the potential will lead to $\sim$ 10 per cent variations in the half mass radius
\citep[][see, e.g., their figures 1]{miholicsetal2014}.

\begin{figure} 
   \includegraphics[width=\columnwidth]{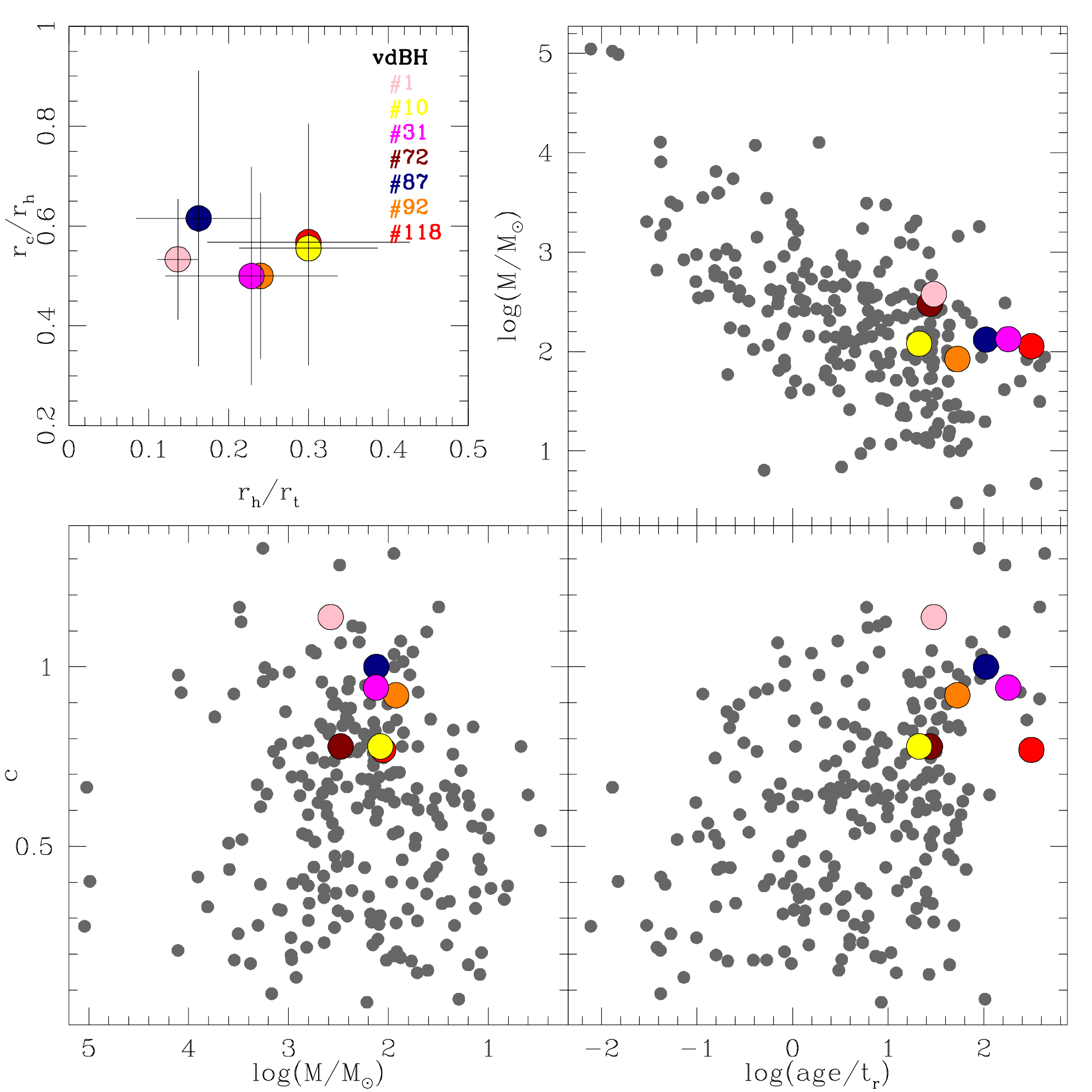}
\caption{Relationships between cluster  core ($r_c$), half-mass ($r_h$) and tidal
($r_t$) radii, concentration parameter ($c$), mass, age and relaxation time ($t_r$).}
Grey dots correspond to 236 star clusters with homogeneous estimations of
masses and radii derived by \citet{piskunovetal2007}.
\label{fig:fig14}
\end{figure}

The age/$t_r$ ratio is a
good indicator of the internal dynamical evolution, since it gives the number of
times the characteristic time-scale to reach some level of energy equipartition 
\citep{bm98} has been surpassed. Star clusters with large age/$t_r$ ratios have reached a
higher degree of relaxation and hence are dynamically more evolved. As 
Fig.~\ref{fig:fig14} shows, our clusters appear to cover a wide range of age/$t_r$ 
ratios, from $\sim$ 20 up to 320, all of them suggesting that the clusters
have had enough time to evolve dynamically. In the figure we included in grey colour 236 
open clusters analysed by \citet{piskunovetal2007}, who derived from them homogeneous
scales of radii and masses. They derived core and tidal radii for their cluster sample,
from which we calculated the half-mass radii and, with their clusters masses and eq. 12, relaxation times, by assuming that the cluster stellar density
profiles can be indistinguishably reproduced by King and Plummer models.
Their cluster sample are mostly distributed inside 
a circle of $\sim$ 1 kpc from the Sun. As can be seen, our clusters cover the most
evolved limit of the age/$t_r$ distribution (right-hand panels).

Since dynamical evolution implies the loss of stars (mass loss), we expect some trend of
the present-day cluster mass with the age/$t_r$ ratio. This is confirmed
in the top-right panel of Fig.~\ref{fig:fig14}, where the larger the present-day
mass the less the dynamical evolution of a cluster in the solar neighbourhood,
with a noticeable scatter. The studied clusters appear to have relatively large masses
for their particular internal dynamical states. Curiously, selection against poor 
and old clusters
could suggest the beggining of cluster dissolution, with some exceptions.

\citet{trentietal2010} presented a unified picture for the evolution of star clusters 
on the two-body relaxation timescale from direct N-body simulations of star clusters
in a tidal field. Their treatment of the stellar evolution is based on the approximation
that most of the relevant stellar evolution occurs on a timescale shorter than a 
relaxation time, when the most massive stars lose a significant fraction of mass and consequently  contribute to a global expansion of the system. Later in the
life of a star cluster, two-body relaxation tends to erase the memory of the initial 
density profile and concentration. They found that the
structure of the system, as measured by the core to half mass radius ratio, the
concentration parameter $c$= log($r_t/r_c$), among others, 
evolve toward a universal state, which is set by the efficiency of heating on
the visible population of stars induced by dynamical interactions in the core of
the system. In the bottom panels of Fig.~\ref{fig:fig14} we plotted the
dependence of the concentration parameter $c$ with the cluster mass and the
age/$t_r$ ratio, respectively. They show that our dynamically evolved clusters are 
within those with relatively high $c$ values, and that star clusters tend to 
initially start their dynamical evolution with relatively small concentration 
parameters. Likewise, star clusters in an advanced dynamical state can also have relatively
lower $c$ values due to their smaller masses.
 
According to \citet[][see, e.g., their figure 33.2]{hh03} a star cluster dynamically
evolving with its tidal radius filled, moves in the $r_c/r_h$ vs $r_h/r_t$ plane 
parallel to the $r_c/r_h$ axis ($r_h/r_t$ $\sim$ 0.21) toward low values due to 
violent relaxation in the cluster core region followed by two-body relaxation,
mass segregation, and finally core-collapse. Top-left panel in Fig.~\ref{fig:fig14}
shows that the studied clusters have  $r_h/r_t$ 
ratios in fairly agreement, within the uncertainties, with that of tidally 
filled clusters, and $r_c/r_h$ ratios suggesting mass segregation in
their core regions.

\begin{figure} 
   \includegraphics[width=\columnwidth]{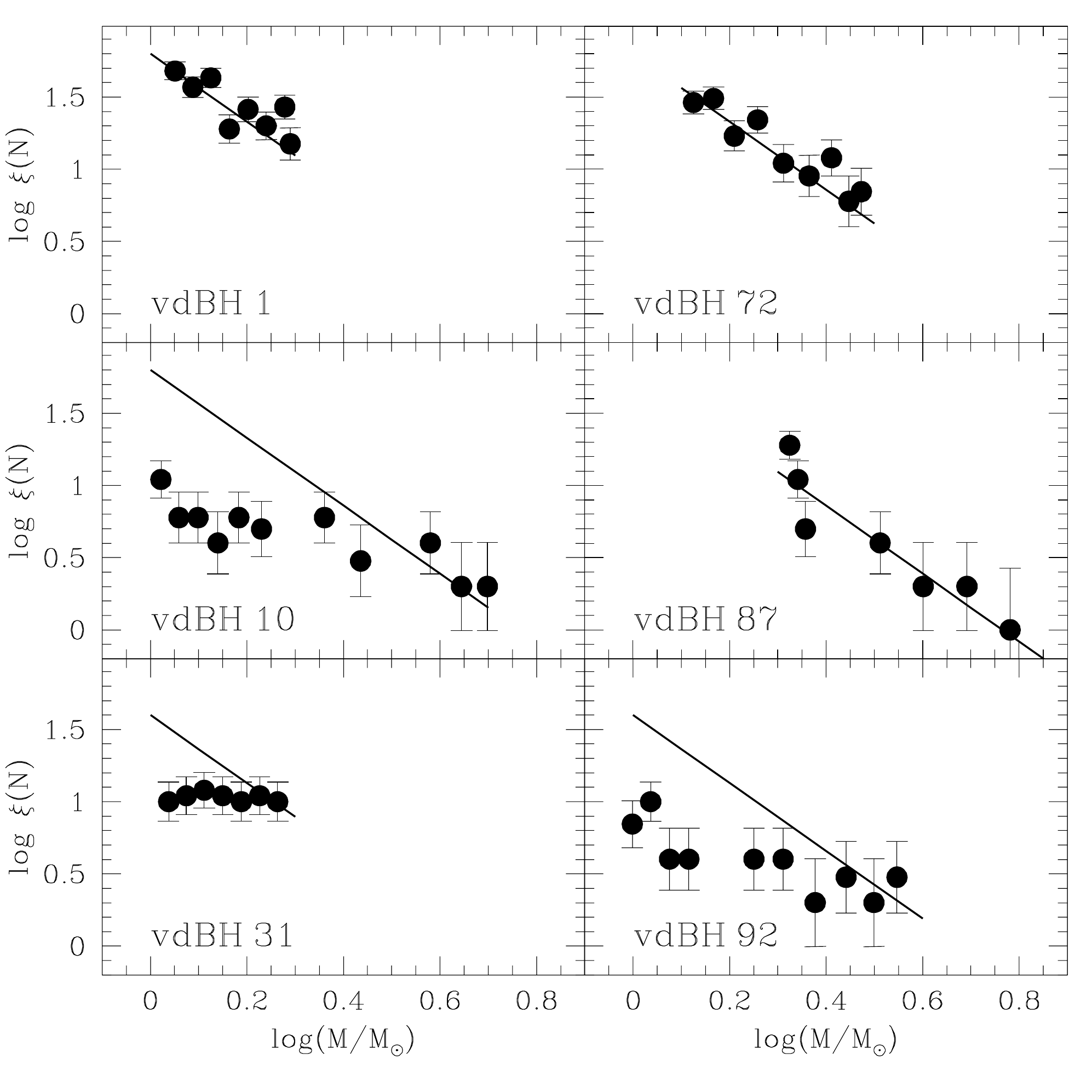}
\caption{Mass function for clusters in our sample. The \citet{salpeter55}' relationship for 
stars in the solar neighbourhood is superimposed.}
\label{fig:fig15}
\end{figure}

We finally built the cluster mass functions (MFs) employing the masses of
stars with photometric memberships $P \ge$ 75\%. The resulting MFs are
shown in Fig.~\ref{fig:fig15} where the errorbars come from applying Poisson statistics. 
We did not include vdBH\,118 because of incompleteness
in our photometry for its MS. For comparison porpuses we superimposed the relationship
given by \citet[][slope = -2.35]{salpeter55} for the stars in the solar neighbourhood.
As can be seen, despite their advanced state of dynamical evolution, vdBH\,1, 72 and 87 
still keep their MFs close to that of Salpeter's law, while the remaining clusters 
(vdBH\,10, 31 and 92) show MFs which depart from it for smaller masses toward a
relation with a smaller slope \citep[see, e.g.][]{limetal2015,santosetal2016}. Their total 
masses, structural parameters and age/$t_r$ ratios do not suggest any hint to explain such 
differences, and prevent us to draw any conclusion from such a small cluster sample.

\section{conclusions}

We present results of seven star clusters identified
by \citet{vdbh75}, namely, vdBH\,1, 10, 31, 72, 87, 92, and 118, for which particular 
attention was not given until now. The clusters were observed 
through the Johnson $UBV$,  Kron-Cousins $RI$ and Washington $C$ filters;  four of 
them (vdBH\,72, 87, 92 and 118) are photometrically study for the first time.

The multi-band photometric data sets were used to trace the cluster stellar
density radial profiles and to build CMDs and CC diagrams, from which we estimate their
structural parameters and fundamental astrophysical properties. Their radial profiles 
were built from star counts carried out throughout the observed
fields using the final photometric catalogues. We derived the cluster radii
from a careful placement of the background levels
and fitted King and Plummer models to derive cluster core, half-mass and tidal radii. 
We then applied a subtraction procedure developed by \citet{pb12} to statistically clean the 
star cluster CMDs and CC diagrams from field star contamination in order to disentangle 
star cluster  features from those belonging to their surrounding fields. The employed 
technique makes use of variable cells in order to reproduce the field CMD as closely 
as possible. 

The availability of  three CC diagrams and six CMDs covering wavelengths from 
the blue up to the near-infrarred allowed us to derive reliable ages, reddenings and 
distances for the studied clusters. We exploited such a wealth in combination with
a new age-metallicity diagnostic diagram for the Washington system and theoretical
isochrones computed by \citet{betal12} to find out that the clusters in our sample
cover a wide age range, from $\sim$ 60 Myr up to 2.8 Gyr, are of relatively small
size ($\sim$ 1 $-$ 6 pc) and are placed at distances from the Sun which vary between
1.8 and 6.3 kpc, respectively. They are located between the Carina and Perseus spiral arms,
with the exception of vdBH\,118, which is closer to the Galactic centre than the
Carina arm.  It also belongs to the sample of the farthest known clusters placed
between the Carina and Crux spiral arms.

We estimated lower limits for the cluster present-day masses as well as half-mass 
relataxion times. Their resulting values suggest that the studied clusters have
advanced states of their internal dynamical evolution, i.e., their ages are many times
the relaxation times ($\sim$ 20 $-$ 320). When combined with the obtained structural
parameters, we found that the clusters are possible in the phase typical of those
tidally filled with mass segregation in their core regions. 

We compared the cluster
masses, concentration parameters and age/$t_r$ ratios with those for 236 clusters located
in the solar neighbourhood. The seven vdBH clusters are within more massive
($\sim$ 80 $-$ 380$\msun$), with higher $c$ values, and dynamically evolved  
ones. There is also a broad correlation between the cluster massses and the
dynamical state, in the sense that the more the advanced  the internal dynamical
evolution, the less the cluster mass. The apparent decrease of clusters
with masses smaller than $\sim$ 200$\msun$ and ages $\sim$ 100 times older than their 
relaxation times could suggest the beggining of cluster dissolution, with some
exceptions. Finally, we found that the MFs of vdBH\,1, 72 and 87 follow approximately the 
\citet{salpeter55}'s relationship, while vdBH\,10, 31 and 92 seem to show shallower
MFs for the lower mass regime.

\section*{Acknowledgements}
We thank the anonymous referee whose thorough comments and suggestions
allowed us to improve the manuscript.




\bibliographystyle{mnras}

\begin{thebibliography}{}
\makeatletter
\relax
\def\mn@urlcharsother{\let\do\@makeother \do\$\do\&\do\#\do\^\do\_\do\%\do\~}
\def\mn@doi{\begingroup\mn@urlcharsother \@ifnextchar [ {\mn@doi@}
  {\mn@doi@[]}}
\def\mn@doi@[#1]#2{\def\@tempa{#1}\ifx\@tempa\@empty \href
  {http://dx.doi.org/#2} {doi:#2}\else \href {http://dx.doi.org/#2} {#1}\fi
  \endgroup}
\def\mn@eprint#1#2{\mn@eprint@#1:#2::\@nil}
\def\mn@eprint@arXiv#1{\href {http://arxiv.org/abs/#1} {{\tt arXiv:#1}}}
\def\mn@eprint@dblp#1{\href {http://dblp.uni-trier.de/rec/bibtex/#1.xml}
  {dblp:#1}}
\def\mn@eprint@#1:#2:#3:#4\@nil{\def\@tempa {#1}\def\@tempb {#2}\def\@tempc
  {#3}\ifx \@tempc \@empty \let \@tempc \@tempb \let \@tempb \@tempa \fi \ifx
  \@tempb \@empty \def\@tempb {arXiv}\fi \@ifundefined
  {mn@eprint@\@tempb}{\@tempb:\@tempc}{\expandafter \expandafter \csname
  mn@eprint@\@tempb\endcsname \expandafter{\@tempc}}}

\bibitem[\protect\citeauthoryear{{Binney} \& {Merrifield}}{{Binney} \&
  {Merrifield}}{1998}]{bm98}
{Binney} J.,  {Merrifield} M.,  1998, {Galactic Astronomy}

\bibitem[\protect\citeauthoryear{{Bonatto} \& {Bica}}{{Bonatto} \&
  {Bica}}{2010}]{bb10}
{Bonatto} C.,  {Bica} E.,  2010, \mn@doi [\mnras]
  {10.1111/j.1365-2966.2010.17000.x}, \href
  {http://adsabs.harvard.edu/abs/2010MNRAS.407.1728B} {407, 1728}

\bibitem[\protect\citeauthoryear{{Bressan}, {Marigo}, {Girardi}, {Salasnich},
  {Dal Cero}, {Rubele}  \& {Nanni}}{{Bressan} et~al.}{2012}]{betal12}
{Bressan} A.,  {Marigo} P.,  {Girardi} L.,  {Salasnich} B.,  {Dal Cero} C.,
  {Rubele} S.,   {Nanni} A.,  2012, \mn@doi [\mnras]
  {10.1111/j.1365-2966.2012.21948.x}, 427, 127

\bibitem[\protect\citeauthoryear{{Cardelli}, {Clayton}  \& {Mathis}}{{Cardelli}
  et~al.}{1989}]{cetal89}
{Cardelli} J.~A.,  {Clayton} G.~C.,   {Mathis} J.~S.,  1989, \mn@doi [\apj]
  {10.1086/167900}, 345, 245

\bibitem[\protect\citeauthoryear{{Carraro}, {Beletsky}  \& {Marconi}}{{Carraro}
  et~al.}{2013}]{cetal13}
{Carraro} G.,  {Beletsky} Y.,   {Marconi} G.,  2013, \mn@doi [\mnras]
  {10.1093/mnras/sts038}, 428, 502

\bibitem[\protect\citeauthoryear{{Dias}, {Alessi}, {Moitinho}  \&
  {L{\'e}pine}}{{Dias} et~al.}{2002}]{detal02}
{Dias} W.~S.,  {Alessi} B.~S.,  {Moitinho} A.,   {L{\'e}pine} J.~R.~D.,  2002,
  \mn@doi [\aap] {10.1051/0004-6361:20020668}, 389, 871

\bibitem[\protect\citeauthoryear{{Drimmel} \& {Spergel}}{{Drimmel} \&
  {Spergel}}{2001}]{ds2001}
{Drimmel} R.,  {Spergel} D.~N.,  2001, \mn@doi [\apj] {10.1086/321556}, \href
  {http://adsabs.harvard.edu/abs/2001ApJ...556..181D} {556, 181}

\bibitem[\protect\citeauthoryear{{Geisler}}{{Geisler}}{1996}]{g96}
{Geisler} D.,  1996, \mn@doi [\aj] {10.1086/117799}, 111, 480

\bibitem[\protect\citeauthoryear{{Giorgi}, {Solivella}, {Perren}  \&
  {V{\'a}zquez}}{{Giorgi} et~al.}{2015}]{giorgietal2015}
{Giorgi} E.~E.,  {Solivella} G.~R.,  {Perren} G.~I.,   {V{\'a}zquez} R.~A.,
  2015, \mn@doi [\na] {10.1016/j.newast.2015.04.004}, \href
  {http://adsabs.harvard.edu/abs/2015NewA...40...87G} {40, 87}

\bibitem[\protect\citeauthoryear{{Han}, {Curtis}  \& {Wright}}{{Han}
  et~al.}{2016}]{hanetal2016}
{Han} E.,  {Curtis} J.~L.,   {Wright} J.~T.,  2016, preprint, \href
  {http://adsabs.harvard.edu/abs/2016arXiv160505330H} {} (\mn@eprint {arXiv}
  {1605.05330})

\bibitem[\protect\citeauthoryear{{Heggie} \& {Hut}}{{Heggie} \&
  {Hut}}{2003}]{hh03}
{Heggie} D.,  {Hut} P.,  2003, {The Gravitational Million-Body Problem: A
  Multidisciplinary Approach to Star Cluster Dynamics}

\bibitem[\protect\citeauthoryear{{Heiter}, {Soubiran}, {Netopil}  \&
  {Paunzen}}{{Heiter} et~al.}{2014}]{hetal14}
{Heiter} U.,  {Soubiran} C.,  {Netopil} M.,   {Paunzen} E.,  2014, \mn@doi
  [\aap] {10.1051/0004-6361/201322559}, 561, A93

\bibitem[\protect\citeauthoryear{{Hiltner} \& {Johnson}}{{Hiltner} \&
  {Johnson}}{1956}]{hj56}
{Hiltner} W.~A.,  {Johnson} H.~L.,  1956, \mn@doi [\apj] {10.1086/146231},
  \href {http://adsabs.harvard.edu/abs/1956ApJ...124..367H} {124, 367}

\bibitem[\protect\citeauthoryear{{Hou} \& {Han}}{{Hou} \& {Han}}{2014}]{hh2014}
{Hou} L.~G.,  {Han} J.~L.,  2014, \mn@doi [\aap] {10.1051/0004-6361/201424039},
  \href {http://adsabs.harvard.edu/abs/2014A%26A...569A.125H} {569, A125}

\bibitem[\protect\citeauthoryear{{King}}{{King}}{1962}]{king62}
{King} I.,  1962, \mn@doi [\aj] {10.1086/108756}, 67, 471

\bibitem[\protect\citeauthoryear{{Landolt}}{{Landolt}}{1992}]{l92}
{Landolt} A.~U.,  1992, \mn@doi [\aj] {10.1086/116242}, \href
  {http://adsabs.harvard.edu/abs/1992AJ....104..340L} {104, 340}

\bibitem[\protect\citeauthoryear{{Lim}, {Sung}, {Hur}  \& {Park}}{{Lim}
  et~al.}{2015}]{limetal2015}
{Lim} B.,  {Sung} H.,  {Hur} H.,   {Park} B.-G.,  2015, preprint, \href
  {http://adsabs.harvard.edu/abs/2015arXiv151101118L} {} (\mn@eprint {arXiv}
  {1511.01118})

\bibitem[\protect\citeauthoryear{{Miholics}, {Webb}  \& {Sills}}{{Miholics}
  et~al.}{2014}]{miholicsetal2014}
{Miholics} M.,  {Webb} J.~J.,   {Sills} A.,  2014, \mn@doi [\mnras]
  {10.1093/mnras/stu1951}, \href
  {http://adsabs.harvard.edu/abs/2014MNRAS.445.2872M} {445, 2872}

\bibitem[\protect\citeauthoryear{{Moitinho}, {V{\'a}zquez}, {Carraro}, {Baume},
  {Giorgi}  \& {Lyra}}{{Moitinho} et~al.}{2006}]{moitinhoetal2006}
{Moitinho} A.,  {V{\'a}zquez} R.~A.,  {Carraro} G.,  {Baume} G.,  {Giorgi}
  E.~E.,   {Lyra} W.,  2006, \mn@doi [\mnras]
  {10.1111/j.1745-3933.2006.00163.x}, \href
  {http://adsabs.harvard.edu/abs/2006MNRAS.368L..77M} {368, L77}

\bibitem[\protect\citeauthoryear{{Paunzen}, {Heiter}, {Netopil}  \&
  {Soubiran}}{{Paunzen} et~al.}{2010}]{paunzeretal2010}
{Paunzen} E.,  {Heiter} U.,  {Netopil} M.,   {Soubiran} C.,  2010, \mn@doi
  [\aap] {10.1051/0004-6361/201014131}, \href
  {http://adsabs.harvard.edu/abs/2010A%26A...517A..32P} {517, A32}

\bibitem[\protect\citeauthoryear{{Piatti} \& {Bica}}{{Piatti} \&
  {Bica}}{2012}]{pb12}
{Piatti} A.~E.,  {Bica} E.,  2012, \mn@doi [\mnras]
  {10.1111/j.1365-2966.2012.21694.x}, 425, 3085

\bibitem[\protect\citeauthoryear{{Piatti} \& {Perren}}{{Piatti} \&
  {Perren}}{2015}]{pp15}
{Piatti} A.~E.,  {Perren} G.~I.,  2015, \mn@doi [\mnras]
  {10.1093/mnras/stv861}, \href
  {http://adsabs.harvard.edu/abs/2015MNRAS.450.3771P} {450, 3771}

\bibitem[\protect\citeauthoryear{{Piskunov}, {Schilbach}, {Kharchenko},
  {R{\"o}ser}  \& {Scholz}}{{Piskunov} et~al.}{2007}]{piskunovetal2007}
{Piskunov} A.~E.,  {Schilbach} E.,  {Kharchenko} N.~V.,  {R{\"o}ser} S.,
  {Scholz} R.-D.,  2007, \mn@doi [\aap] {10.1051/0004-6361:20077073}, \href
  {http://adsabs.harvard.edu/abs/2007A%26A...468..151P} {468, 151}

\bibitem[\protect\citeauthoryear{{Salpeter}}{{Salpeter}}{1955}]{salpeter55}
{Salpeter} E.~E.,  1955, \mn@doi [\apj] {10.1086/145971}, \href
  {http://adsabs.harvard.edu/abs/1955ApJ...121..161S} {121, 161}

\bibitem[\protect\citeauthoryear{{Santos}, {Roman-Lopes}, {Carrasco}, {Maia}
  \& {Neichel}}{{Santos} et~al.}{2016}]{santosetal2016}
{Santos} Jr. J.~F.~C.,  {Roman-Lopes} A.,  {Carrasco} E.~R.,  {Maia} F.~F.~S.,
   {Neichel} B.,  2016, \mn@doi [\mnras] {10.1093/mnras/stv2731}, \href
  {http://adsabs.harvard.edu/abs/2016MNRAS.456.2126S} {456, 2126}

\bibitem[\protect\citeauthoryear{{Skrutskie} et~al.,}{{Skrutskie}
  et~al.}{2006}]{skrutskieetal2006}
{Skrutskie} M.~F.,  et~al., 2006, \mn@doi [\aj] {10.1086/498708}, \href
  {http://adsabs.harvard.edu/abs/2006AJ....131.1163S} {131, 1163}

\bibitem[\protect\citeauthoryear{{Spitzer} \& {Hart}}{{Spitzer} \&
  {Hart}}{1971}]{sh71}
{Spitzer} Jr. L.,  {Hart} M.~H.,  1971, \mn@doi [\apj] {10.1086/150855}, 164,
  399

\bibitem[\protect\citeauthoryear{{Stetson}, {Davis}  \& {Crabtree}}{{Stetson}
  et~al.}{1990}]{setal90}
{Stetson} P.~B.,  {Davis} L.~E.,   {Crabtree} D.~R.,  1990, in {Jacoby} G.~H.,
  ed.,  Astronomical Society of the Pacific Conference Series Vol. 8, CCDs in
  astronomy. pp 289--304

\bibitem[\protect\citeauthoryear{{Trenti}, {Vesperini}  \& {Pasquato}}{{Trenti}
  et~al.}{2010}]{trentietal2010}
{Trenti} M.,  {Vesperini} E.,   {Pasquato} M.,  2010, \mn@doi [\apj]
  {10.1088/0004-637X/708/2/1598}, \href
  {http://adsabs.harvard.edu/abs/2010ApJ...708.1598T} {708, 1598}

\bibitem[\protect\citeauthoryear{{V{\'a}zquez}, {May}, {Carraro}, {Bronfman},
  {Moitinho}  \& {Baume}}{{V{\'a}zquez} et~al.}{2008}]{vetal08}
{V{\'a}zquez} R.~A.,  {May} J.,  {Carraro} G.,  {Bronfman} L.,  {Moitinho} A.,
   {Baume} G.,  2008, \mn@doi [\apj] {10.1086/524003}, 672, 930

\bibitem[\protect\citeauthoryear{{van den Bergh} \& {Hagen}}{{van den Bergh} \&
  {Hagen}}{1975}]{vdbh75}
{van den Bergh} S.,  {Hagen} G.~L.,  1975, \mn@doi [\aj] {10.1086/111707},
  \href {http://adsabs.harvard.edu/abs/1975AJ.....80...11V} {80, 11}

\makeatother
\end{thebibliography}

\input{paper.bbl}


\setcounter{figure}{4}
\begin{landscape}
\begin{figure} 
\caption{CMDs and CC diagrams for stars measured in the field of vdBH\,1. Symbols
are as in Fig.~\ref{fig:fig4}. We overplotted the isochrone which best matches the
cluster features (see text for details).}
   \label{fig:fig5}
\end{figure}
\end{landscape}

\begin{landscape}
\begin{figure} 
\caption{CMDs and CC diagrams for stars measured in the field of vdBH\,10. Symbols
are as in Fig.~\ref{fig:fig4}. We overplotted the isochrone which best matches the
cluster features (see text for details).}
\label{fig:fig6}
\end{figure}
\end{landscape}

\begin{landscape}
\begin{figure} 
\caption{CMDs and CC diagrams for stars measured in the field of vdBH\,31. Symbols
are as in Fig.~\ref{fig:fig4}. We overplotted the isochrone which best matches the
cluster features (see text for details).}
\label{fig:fig7}
\end{figure}
\end{landscape}

\begin{landscape}
\begin{figure} 
\caption{CMDs and CC diagrams for stars measured in the field of vdBH\,72. Symbols
are as in Fig.~\ref{fig:fig4}. We overplotted the isochrone which best matches the
cluster features (see text for details).}
\label{fig:fig8}
\end{figure}
\end{landscape}

\begin{landscape}
\begin{figure} 
\caption{CMDs and CC diagrams for stars measured in the field of vdBH\,87. Symbols
are as in Fig.~\ref{fig:fig4}. We overplotted the isochrone which best matches the
cluster features (see text for details).}
\label{fig:fig9}
\end{figure}
\end{landscape}

\begin{landscape}
\begin{figure} 
\caption{CMDs and CC diagrams for stars measured in the field of vdBH\,92. Symbols
are as in Fig.~\ref{fig:fig4}. We overplotted the isochrone which best matches the
cluster features (see text for details).}
\label{fig:fig10}
\end{figure}
\end{landscape}

\begin{landscape}
\begin{figure} 
\caption{CMDs and CC diagrams for stars measured in the field of vdBH\,118. Symbols
are as in Fig.~\ref{fig:fig4}. We overplotted the isochrone which best matches the
cluster features (see text for details).}
\label{fig:fig11}
\end{figure}
\end{landscape}






\bsp	
\label{lastpage}


\end{document}